\documentclass[journal=jctcce,manuscript=article,layout=traditional]{achemso}

\setkeys{acs}{articletitle = true} % show reference article titles

\usepackage{amssymb,amsmath,mathrsfs}  % Math typesetting
\usepackage{tikz}
\usepackage{mathtools}		  % for "MoveEqLeft" in aligning equations
\usepackage{color}                                          % Use colors in texts, etc
\usepackage{soul}                                           % for strike-throughs

\title{Absolutely Localized Projection-Based Embedding for Excited States}
\author{Xuelan Wen}
\author{Daniel S. Graham}
\author{Dhabih V. Chulhai}
\affiliation{Department of Chemistry, University of Minnesota. 207 Pleasant St. SE, Minneapolis MN 55455, USA.}
\alsoaffiliation{Current address: Department of Chemistry, University of Indianapolis. 1400 E. Hanna Ave., Indianapolis IN 46227, USA.}
\author{Jason D. Goodpaster}
\email{jgoodpas@umn.edu}
\affiliation{Department of Chemistry, University of Minnesota. 207 Pleasant St. SE, Minneapolis MN 55455, USA.}
\date{\today}

\begin{document}

\maketitle

%%% Abstract [ONLY 75 WORDS (100 words OK though)]
\abstract{
We present a quantum embedding method that allows for the calculation of local excited states embedded in a Kohn-Sham density functional theory (DFT) environment.  Projection-based quantum embedding methodologies provide a rigorous framework for performing DFT-in-DFT and wave function in DFT (WF-in-DFT) calculations.  The use of absolute localization, where the density of each subsystem is expanded in only the basis functions associated with the atoms of that subsystem, provide improved computationally efficiency for WF-in-DFT calculations by reducing the number of orbitals in the WF calculation.  In this work, we extend absolutely localized projection-based quantum embedding to study localized excited states using EOM-CCSD-in-DFT and TDDFT-in-DFT. The embedding results are highly accurate compared to the corresponding canonical EOM-CCSD and TDDFT results on the full system, with TDDFT-in-DFT frequently more accurate than canonical TDDFT.  The absolute localization method is shown to eliminate the spurious low-lying excitation energies for charge transfer states and prevent over delocalization of excited states. Additionally, we attempt to recover the environment response caused by the electronic excitations in the high-level subsystem using different schemes and compare their accuracy.  Finally, we apply this method to the calculation of the excited state energy of green fluorescent protein and show that we systematically converge to the full system results.  Here we demonstrate how this method can be useful in understanding excited states, specifically which chemical moieties polarize to the excitation.   This work shows absolutely localized projection-based quantum embedding can treat local electronic excitations accurately, and make computationally expensive WF methods applicable to systems beyond current computational limits. 
}

\section{Introduction} %=====================================================================

Understanding photo-induced processes in chemical, biological, and material systems necessitates accurate descriptions of electronic excited states.
Linear-response time-dependent density functional theory (LR)-TDDFT\cite{Runge1984-yr,Casida1995-xn}---the excited state extension to ground state density functional theory (DFT)\cite{Hohenberg1964-sd,Kohn1965-bc}---is oftentimes the method of choice for excited states of medium to large molecules (up to a few hundreds of atoms) due to its reasonable cost scaling (formally $\mathscr{O}(N^4)$\cite{Herbert2016,Ding2017}, or $\mathscr{O}(N^3)$\cite{ORourke2015,Zuehlsdorff2015,Wu2011,Gross2012} depending on implementation) and accuracy.
However, implementations of TDDFT rely on approximate exchange-correlation functionals and typically utilize the adiabatic approximation.\cite{Bauernschmitt1996}
In general, TDDFT is found to describe valence excitations quite accurately, however, these approximations lead to well-studied shortcomings of TDDFT in accurately describing Rydberg states, long range charge-transfer excitations, conical intersections, and double excitations\cite{Adamo2013,Dreuw2004}.

Excited states properties may be calculated with high accuracy using \textit{ab initio} correlated wave function (WF) methods.
Equation of motion coupled cluster with singles and doubles (EOM-CCSD)\cite{Stanton1993,Krylov2008-sd} has been shown to be accurate for the excited states of many systems\cite{Schreiber2008}. %\xw{It's interesting that Manby's paper calls EOM-CCSD gold-standard}
However, the computational cost of EOM-CCSD (scaling as $\mathscr{O}(N^6)$) restricts its usage to about 50 or fewer atoms in a moderate basis.
Methods that extend the applicability of EOM-CCSD to larger systems include the use of local orbitals,\cite{Korona2003-ze,Baudin2016-wj,Hofener2017-zw,Dutta2016-bm,Baudin2017-ep} restricted virtual spaces,\cite{Kaliman2017-dl,Epifanovsky2013-zx} and multiscale approaches.\cite{Sneskov2011-pu,Sneskov2011-fr,Hofener2016-vg,Daday2014-ee}
Multiscale approaches treat different regions of the molecule with methods of varying cost and accuracy to reflect their importance.\cite{doi:10.1021/acscatal.6b01387, doi:10.1021/acs.accounts.6b00356}
For example, one may use EOM-CCSD to describe the important region of the molecule while using molecular mechanics (MM), in EOM-CCSD-in-MM methods\cite{Caricato2012-ug,Caricato2012-hz,Caricato2013-eh}, or DFT, in EOM-CCSD-in-DFT methods\cite{Bennie2017}, to describe the remainder of the molecule.
Such approaches take advantage of the high accuracy of EOM-CCSD and the low scaling of MM or DFT to go beyond the size/accuracy limits of any one method alone.

Density functional theory embedding provides a formally exact framework for performing multiscale calculations. This approach has been shown to be accurate for combining density functional theory with wave function theory.
The key choice in density function theory embedding is the treatment of the non-additive kinetic energy which can be treated using either approximate kinetic functionals\cite{Wesolowski1993,Wesolowski2015,Jacob2008,doi:10.1021/jp511275e}, or optimized effective potentials\cite{Govind1998,Goodpaster2010,Goodpaster2011,Elliott2010,Huang2011,Fux2010,Huang2018}.
An alternative approach is to employ projection operators to avoid the use of a non-additive kinetic potential\cite{Manby2012,Goodpaster2014,Claudino2019,Tamukong2014-td,doi:10.1063/1.5055942}.  

Projection-based embedding is a density functional theory embedding methodology where one uses projection operators to ensure that the orbitals of two or more regions (or subsystems) of the molecule are mutually orthogonal. This projection operator enforces the Pauli exclusion principle between subsystems and avoids the need to use non-additive kinetic energy functionals. 
Two such projection operators have since been used: (1) the parameter dependent operator (hereinafter the $\mu$ operator);\cite{Manby2012,Goodpaster2014} and (2) the Huzinaga operator.\cite{Huzinaga1971-wk,Hegely2016-cy}
%\xw{repeat with the content later:Bennie and coworkers have presented an embedded EOM-CCSD-in-DFT method based on the $\mu$ operator that reproduces canonical EOM-CCSD results at a reduced cost for solute-in-solvent systems.\cite{Bennie2017}}
%\xw{It has been explored in EMFT paper. However, the accuracy of this approach to describe excited states of systems divided across covalent bonds remains unexplored.}

Extension of projection-based embedding for excited states has been explored by several groups with applications to EOM-CCSD\cite{Bennie2017}, complete active space self-consistent field (CASSCF),\cite{DeLimaBatista2017} linear response TDDFT (LR-TDDFT),\cite{Chulhai2016-wg,Ding2017, doi:10.1063/1.4807059} real-time TDDFT (RT-TDDFT)\cite{Koh2017}, and density matrix renormalization group (DMRG)\cite{doi:10.1021/acs.jctc.6b00476}.
Of primary concern is the polarization response of the environment subsystem B to an excitation in the high-level subsystem A. Chulhai and Jensen introduced an environment response term derived from the $\mu$ operator.\cite{Chulhai2016-wg}
Bennie and co-workers accounted for this response by systematically including important environment orbitals in the description of the high-level subsystem,\cite{Bennie2017} though this increases the computational cost of the EOM-CCSD calculations.
%\xw{I think the process of finding the important orbitals is more time-consuming than the increased cost on EOM-CCSD}

%xw{move}Furthermore, capturing the polarization response of the environment is an interesting challenge in projection-based EOM-CCSD embedding.
%Bennie and coworkers accounted for this by selectively including more orbitals in the embedded EOM-CCSD calculations,\cite{Bennie2017}  and therefore more efficient strategies are needed.

Chulhai and Goodpaster\cite{Chulhai2017,Chulhai2018} have presented an absolute localization projection-based embedding method based on the Huzinaga operator.\cite{Huzinaga1971-wk,Francisco1992-si}
In the absolute localization method, the embedded wave function region is restricted to the basis functions of the embedded subsystem only, thereby significantly reducing the computational cost.
In this paper, we extend the absolute localization projection-based embedding method to describe excited states using both EOM-CCSD and TDDFT.
We show that this approach is able to accurately describe localized excitations in subsystems that are divided across covalent bonds, eliminate spurious low-lying charge-transfer excitations in TDDFT, and efficiently account for environment polarization.

\section{Theory} %==============================================================

\subsection{Absolute Localization Projection-Based Embedding} \label{subsec:theory_ground} %-----------------

In projection-based embedding methods, the total system is divided \latin{via} the electron density such that for two subsystems 

\begin{equation}
   \gamma_\text{tot}^0 = \gamma_\text{A}^0 + \gamma_\text{B}^0
\end{equation}

\noindent
where $\gamma_\text{A}^0$, $\gamma_\text{B}^0$, and $\gamma_\text{tot}^0$ are the ground state density matrices of subsystems A and B, and the total system, respectively.
The Fock matrix of subsystem A embedded in the environment of B is

\begin{equation}
   \mathbf{f}^\text{A-in-B} = \mathbf{h} + \mathbf{J}[\gamma_\text{A}^0 + \gamma_\text{B}^0] + \mathbf{v}_\text{xc}[\gamma_\text{A}^0 + \gamma_\text{B}^0] + \mathbf{P}^\text{B}
\end{equation}

\noindent
where $\mathbf{h}$ is the total one-electron Hamiltonian that contains the kinetic and nuclear potentials, $\mathbf{J}$ is the Coulomb potential, $\mathbf{v}_\text{xc}$ is the exchange-correlation (XC) potential, and $\mathbf{P}^\text{B}$ is the projection operator that enforces inter-subsystem orbital orthogonality.

The Manby and Miller groups proposed the use of the parameter-dependent level-shift projection operator (hereinafter the $\mu$ operator) as\cite{Manby2012,doi:10.1021/ct5011032,doi:10.1021/acs.jctc.5b00630,doi:10.1021/acs.jctc.6b00685,doi:10.1021/acs.jctc.6b01065,doi:10.1063/1.5050533}

\begin{equation}
   \textbf{P}^\text{B} =  \mu \left( \mathbf{S}^\text{AB} \gamma_\text{B}^0 \mathbf{S}^\text{BA} \right)
   \label{eq:MM-operator}
\end{equation}

\noindent
where $\mu$ is a very large parameter (usually $10^6$), and $\mathbf{S}^\text{AB(BA)}$ are the AO overlap matrices between subsystems A and B. This operator has since seen widespread application and development,\cite{Barnes2013,Goodpaster2014,Tamukong2014-td,Chulhai2015-tq,Bennie2015-ur,Bennie2016-tq,Pennifold2017-ih} including basis set truncation\cite{Barnes2013,Bennie2015-ur,Bennie2016-tq} and extensions to describe excited states.\cite{Chulhai2016-wg,Ding2017,Bennie2017,DeLimaBatista2017,Koh2017}

Alternatively, the Huzinaga operator\cite{Huzinaga1971-wk,Francisco1992-si} may also be used in DFT embedding;\cite{Hegely2016-cy,Chulhai2017,Chulhai2018,Hegely2018-ri} it is defined as

\begin{equation}
   \mathbf{P}^\text{B} = - \frac{1}{2} \left( \mathbf{F}^\text{AB} \gamma_\text{B}^0 \textbf{S}^\text{BA} + \textbf{S}^\text{AB} \gamma_\text{B}^0 \mathbf{F}^\text{BA} \right)
   \label{eq:huzinaga-operator}
\end{equation}

\noindent
where $\mathbf{F}^{AB(BA)}$ are the AB (or BA) block of the total Fock matrix.
The Huzinaga operator yields accurate results when each subsystem density is expanded in only basis functions associated with that subsystem.\cite{Chulhai2017,Chulhai2018}
In these cases, the basis sets of A and B are mutually exclusive and the electrons of A and B are restricted to their respective subsystems only. This strategy, termed ``absolute localization'', has been shown to increase accuracy through error cancellation.\cite{Chulhai2017,Chulhai2018} This allows one to significantly reduce the size of basis sets used to describe subsystems A and B, and therefore significantly reduce the computational cost of the WF calculation.

Standard EOM-CCSD has computational scaling of the order ${O}^2_{A+B}{V}^4_{A+B}$, where $O$ indicates the number of occupied orbitals and $V$ indicates the number of virtual orbitals.
When subsystem density is expanded in the basis functions of the full system (or full-system basis, as shown later), WF-in-DFT calculations scale as $\text{O}^2_{A}{V}^4_{A+B}$.
When the subsystem density is expanded in only the basis functions associate with itself (or subsystem basis), WF-in-DFT calculations scales as $\text{O}^2_{A}{V}^4_{A}$. Therefore, the size of subsystem B could be as large as the computational limit of DFT for EOM-CCSD-in-DFT calculations.

\subsection{Absolutely-Localized Embedding for Excited States} \label{subsec:theory_excited}  %----------------------------

In this paper, we separate the environment polarization into two parts---a ground state polarization and a polarization response---and we introduce two approaches to include the latter in absolute localization embedding for excited states.
For the ground-state polarization, we use the ground state embedding potential---obtained from a ground state absolute localization embedding method as in ref.~\citenum{Chulhai2017}---for excited states calculations.
That is, the embedding potential is built using the DFT ground-state densities of subsystems A and B ($\gamma_\text{A}^0$ and $\gamma_\text{B}^0$) and included through the use of a modified core Hamiltonian ($\mathbf{h}^\text{A-in-B}$) defined as

\begin{equation}
    \mathbf{h}^\text{A-in-B}[\gamma_\text{A}^0,\gamma_\text{B}^0] = \mathbf{h} + \mathbf{J}[\gamma_\text{tot}^0] - \mathbf{J}[\gamma_\text{A}^0] + \mathbf{v}_\text{xc}[\gamma_\text{tot}^0] - \mathbf{v}_\text{xc}[\gamma_\text{A}^0] + \mathbf{P}^\text{B}
\end{equation}

\noindent
This modified core Hamiltonian is then included in all subsequent correlated excited states WF calculations.
However, as previously mentioned, this strategy ignores the polarization response of the environment subsystem B.

In order to account for the polarization response of the environment, we propose two approaches. The first is a state-averaged approach that polarizes subsystem B to an average of the ground state and excited state(s) densities of subsystem A. This state-averaged polarization is included in a self-consistent method as follows (the algorithm is include in Figure~\ref{fig:algorithm}):

\begin{figure}[ht]
    \includegraphics[width=0.5\linewidth]{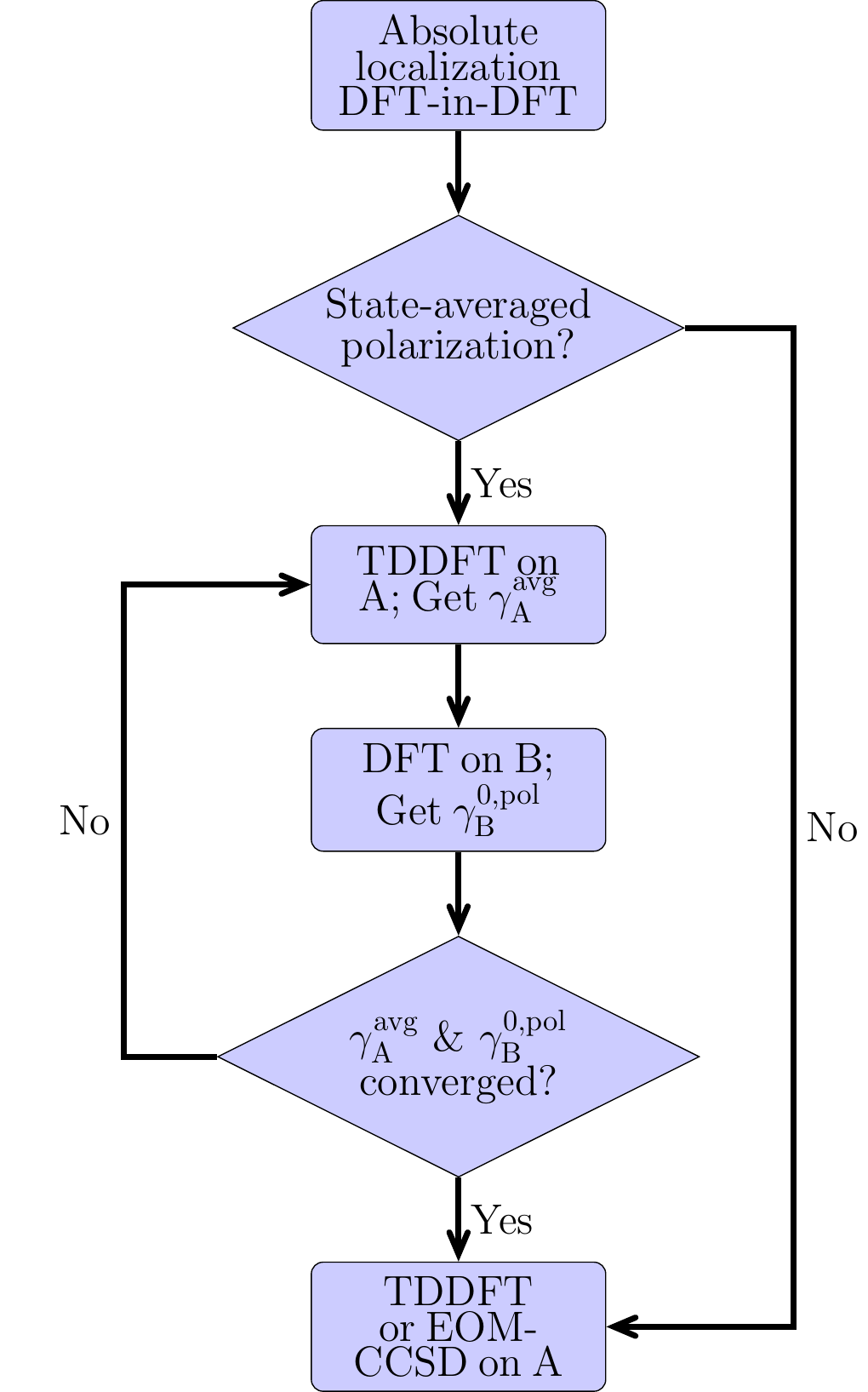}
%  \includestandalone[width=8.0 cm]{flowchart}%     without .tex extension
  % or use \input{mytikz}
  \caption{Algorithm for including state-average polarization response.}
  \label{fig:algorithm}
\end{figure}

(1) Perform absolute localization DFT-in-DFT to obtain self consistent densities $\gamma_\text{A}^0$ and $\gamma_\text{B}^0$, and embedded core Hamiltonian $\mathbf{h}^\text{A-in-B}[\gamma_\text{A}^0,\gamma_\text{B}^0]$.

(2) Perform TDDFT on subsystem A to obtain excited state(s) densities $\gamma_\text{A}^i$, where $i$ denotes the \textit{i}th excited state. The transition density matrix $\gamma_\text{A}^{0 \to i}$ is defined in the eqn. 23 from ref. \citenum{Furche2002}.

(3) Calculate the state-averaged electron density of subsystem A ($\gamma_\text{A}^\text{avg}$). This state-averaged electron density, assuming that we are interested in the first $n$ excited states, is defined as 
\begin{equation}
    \gamma^\text{avg}_\text{A} = \frac{1}{n+1} \sum_{i=0}^n \gamma_\text{A}^i
\end{equation}

(4) Re-optimize the ground state of subsystem B using the new embedded core Hamiltonian $\mathbf{h}^\text{B-in-A}[\gamma_\text{B}^0,\gamma^\text{avg}_\text{A}]$ to obtain a polarized electron density $\gamma^\text{0,pol}_\text{B}$, where 0 denotes the ground state. This step allows subsystem B to be polarized by $\gamma^\text{avg}_\text{A}$.

(5) Re-optimize the ground state of subsystem A with $\mathbf{h}^\text{A-in-B} [\gamma^\text{avg}_\text{A}, \gamma^\text{0,pol}_\text{B}]$ to obtain a polarized $\gamma_\text{A}^\text{0,pol}$. This step allows the electron density in A to be polarized by $\gamma^\text{0,pol}_\text{B}$.

(6) Repeat steps 2--5 using $\gamma_\text{A}^\text{0,pol}$ instead of $\gamma_\text{A}^0$ until both $\gamma^\text{0,pol}_\text{A}$ and $\gamma^\text{0,pol}_\text{B}$ are converged; the convergence criteria is set to $10^{-3}$ for the Frobenius norm of the change in the density matrices in this paper. 

(7) Perform excited states WF (or TDDFT) calculations using the embedded core Hamiltonian $\mathbf{h}[\gamma_\text{A}^\text{0,pol}, \gamma_\text{B}^\text{0,pol}]$ with the converged $\gamma_\text{A}^\text{0,pol}$ and $\gamma_\text{B}^\text{0,pol}$.

The second approach to account for the environment polarization response, only applicable to embedded excited states WF methods, is by using a correction term obtained by TDDFT on the full system, defined as

\begin{equation}
   \begin{aligned}
\MoveEqLeft \tilde{\omega}_\text{WF-in-DFT}[\Psi^\text{A};\gamma_\text{A}^0, \gamma_\text{B}^0,\gamma_\text{KS-DFT}^0]   \\
 & =\omega_\text{TDDFT}[\gamma_\text{KS-DFT}^0]  - \omega_\text{TDDFT-in-DFT}[\gamma_\text{A}^0, \gamma_\text{B}^0]  \\
 & +\omega_\text{WF-in-DFT}[\Psi^\text{A};\gamma_\text{A}^0, \gamma_\text{B}^0] 
   \end{aligned}
   \label{eq:corrected_wf_energy}
\end{equation}

\noindent
where $\tilde{\omega}_\text{WF-in-DFT}$ is the corrected embedded WF excitation energy, $\omega_\text{WF-in-DFT}$ is the embedded WF excitation energy (with ground-state polarization or state-average polarization), $\omega_\text{TDDFT-in-DFT}$ is the embedded TDDFT excitation energy, and $\omega_\text{TDDFT}$ is the canonical TDDFT excitation on the full system with KS-DFT ground density $\gamma_\text{KS-DFT}^0$.
This correction term ($ \omega_\text{TDDFT} - \omega_\text{TDDFT-in-DFT}$) is meant to account for missing environment polarization response for WF methods at the TDDFT level and is further explored in Section~\ref{subsec:acrolein}.

\section{Computational details}%=================================================================

All embedded TDDFT calculations were perform in our Quantum Solid State and Molecular Embedding (QSoME) code\cite{qsome2019}, which utilizes the Python-based Simulations of Chemistry Framework (PySCF).\cite{Sun2018}
All full-system and embedded EOM-CCSD calculations were performed using Molpro 2015.1\cite{MOLPRO,MOLPRO-WIREs}.
For embedded EOM-CCSD calculations, the embedded core Hamiltonian $\mathbf{h}^\text{A-in-B}$ is calculated and then exported from QSoME to Molpro for the EOM-CCSD calculations only.
All full-system TDDFT calculations and Natural transition orbitals (NTOs) were performed using Gaussian 16 program\cite{g16}. 

The geometries of long-chain hydrocarbons with different functional groups (Section~\ref{subsec:ten-carbon}) were obtained using the M06 functional\cite{Zhao2008-hk} and the cc-pVTZ basis set\cite{Dunning1989-yx} using Gaussian 16 program\cite{g16}; all other calculations used the cc-pVDZ basis set.\cite{Dunning1989-yx}
Geometries for acrolein in water in Section~\ref{subsec:adenine} were generated using molecular dynamics, and the aug-cc-pVDZ basis set\cite{Kendall1992} was used for all embedding calculations.
The geometry for acrolein in two water molecules in Section~\ref{subsec:acrolein} was taken from ref~\citenum{Bennie2017}, and the aug-cc-pVDZ basis set was used for all calculations;
the state-average density includes the ground state and the first excited state.
Geometries for \textit{cis}-7HQ (Section~\ref{subsec:7HQ}) were taken from ref~\citenum{Fradelos2011} and the aug-cc-pVDZ basis set was used for all calculations.
The geometry for GFP (Section~\ref{subsec:GFP}) was taken from ref~\citenum{Kaila2013}; the B3LYP functional\cite{Becke1993-js,Stephens1994-da} and def2-TZVP basis set\cite{Weigend2005-fr} were used for all calculations.

\section{Results and Discussion}%========================================================

%\dvcs{Section 4.1 compares the results of absolute localization and $\mu$ operator with supermolecular basis on long-chain organic molecules. Section 4.2 compares the results of absolute localization and $\mu$ operator with monomolecular basis on acrolein solvated in two water molecules. It shows that using absolute localization removes the need for sophisticated strategies for choosing occupied and virtual space. Section 4.3 shows the ability of absolute localization to remove spurious low-lying intramolecular charge-transfer states while treat localized excitations accurately using adenine dimer, a pi-stacked systems. Section 4.4 shows EOM-CCSD-in-DFT and TDDFT-in-DFT results using the 2 algorithms in section 2 for 7HQ chromophore. Section 4.5 applies absolute localization to a 161-atom cluster model of green fluorescent protein, demonstrating its ability to speed up and push up the limit for calculations of large systems.}

\subsection{Choice of Basis for the Subsystem} \label{subsec:ten-carbon} %-------------------

%The NTO pair with the biggest eigenvector. The isovalue for NTOs is set to 0.02.} \dvcc{Moved to Computational Details}

%In projection-based embedding, one uses level shift projection operators to ensure that the occupied orbitals of each subsystem are mutually orthogonal. In their seminal paper, Manby, Miller and co-workers\cite{Manby2012} used a parameter-dependent operator, shown in eqn.~\ref{eq:MM-operator}, that is based on earlier methods,\cite{Phillips1959-ti,Lykos1956-mu,Stoll2005-qy,Mata2008-if,Henderson2006-zc,Gomes2012-pa}  to achieve mutual orthogonality. There has since been many extensions of this $\mu$ projection-based technique,\cite{Barnes2013,Goodpaster2014,Tamukong2014-td,Chulhai2015-tq,Bennie2015-ur,Bennie2016-tq,Pennifold2017-ih} including descriptions of excited states --- such as applications to EOM-CCSD,\cite{Bennie2017} complete active space self-consistent field (CASSCF),\cite{DeLimaBatista2017} linear response TDDFT (LR-TDDFT),\cite{Chulhai2016-wg,Ding2017} and real-time TDDFT (RT-TDDFT).\cite{Koh2017}

First, we will focus on the choice of basis functions to use in embedding calculations for excited states embedded correlated wave function methods --- namely, equation of motion coupled-cluster with single and double excitations (EOM-CCSD).\cite{Stanton1993} EOM-CCSD scales as $N_\text{occ}^2N_\text{vir}^4$, where $N_\text{occ}$ and $N_\text{vir}$ are the number of occupied and virtual orbitals, respectively. Therefore, embedding calculations, even if performed in the full basis of the full system still benefit since the number of occupied orbitals in the high-level subsystem $N^\text{A}_\text{occ}$ may be significantly reduced ($N^\text{A}_\text{occ} \ll N_\text{occ}$). One may further improve the efficiency of such methods by employing basis set truncation schemes\cite{Barnes2013,Bennie2015-ur,Bennie2016-tq}, thereby reducing the available virtual orbitals of the subsequent WF calculation ($N^\text{trunc}_\text{vir} < N_\text{vir}$).\cite{Barnes2013,Bennie2015-ur,Bennie2017}

%However, $\mu$ operator does not commute with the Fock operator in a truncated basis.\cite{Hegely2016-cy,Chulhai2017} %This may lead to convergence issues in (severely) truncated basis sets.\cite{Chulhai2017} 
%Because of this, K\'{a}llay and co-workers\cite{Hegely2016-cy,Hegely2018-ri} used the Huzinaga operator,\cite{Huzinaga1971-wk,Francisco1992-si} which does commute with the Fock operator.

With the Huzinaga operator, we have shown that we can truncate the basis to only include those that describe the atoms in subsystem A.\cite{Chulhai2017,Chulhai2018} This truncation restricts the electrons of subsystem A to that subsystem only, and we label this method as ``absolute localization.'' For ground state properties of both molecular and extended materials, this absolute localization strategy for WF-in-DFT provides similar or improved accuracy to other embedding schemes that also use projection operators at a significantly reduced cost. In this subsection, we will continue to explore the performance of Huzinaga and absolute localization embedding strategies for embedded excited states WF-in-DFT.

\begin{figure}
    \centering
    \includegraphics[width=\linewidth]{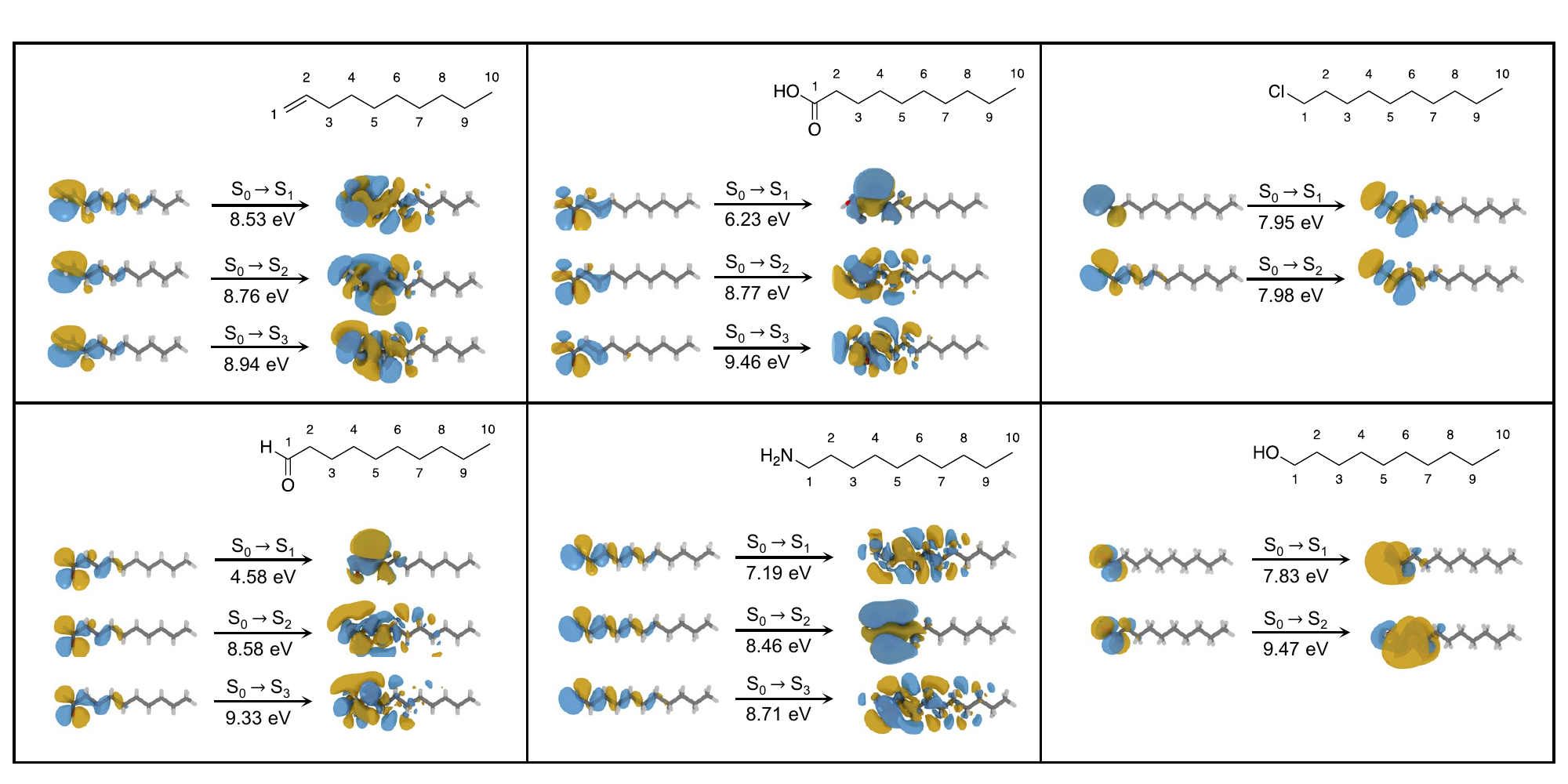}
    \caption{Natural transition orbitals (NTOs) of the lowest three singlet excitations of decene, decanal, decanol, decanamine, chlorodecane and decanoic acid. Shown on highest occupied NTO (HONTO) and lowest unoccupied NTO (LUNTO).}
    \label{fig:NTOs}
\end{figure}

In order to compare these embedding strategies, we will look at their accuracy and efficiency at correctly reproducing the localized excited states in six long-chain organic molecules. We chose the three lowest singlet excitations in decanoic acid, decan-1-amine, dec-1-ene, and decanal, and the two lowest excitations in 1-chlorodecane and decan-1-ol.
The NTOs for these excitations are shown in Figure~\ref{fig:NTOs}. As seen in this figure, these excitations are localized on the functional groups and up to five carbon atoms of the alkane chains.
The third excited states in 1-chlorodecane and decan-1-ol are delocalized (or global) excitations; these results are included in the Supporting Information Figure~S1 and Figure~S2.
We chose long-chain molecules in order to systematically increase the size of the high-level subsystem while testing the accuracy of the methods. Note that, in each case, we will be cutting across covalent bonds. Ding and co-workers examined similar long-chain organic molecules, albeit for embedded TDDFT calculations, in order to show that the $\mu$ strategy is systematically improvable.\cite{Ding2017}

\begin{figure}[!ht]
    \includegraphics[width=0.6\linewidth]{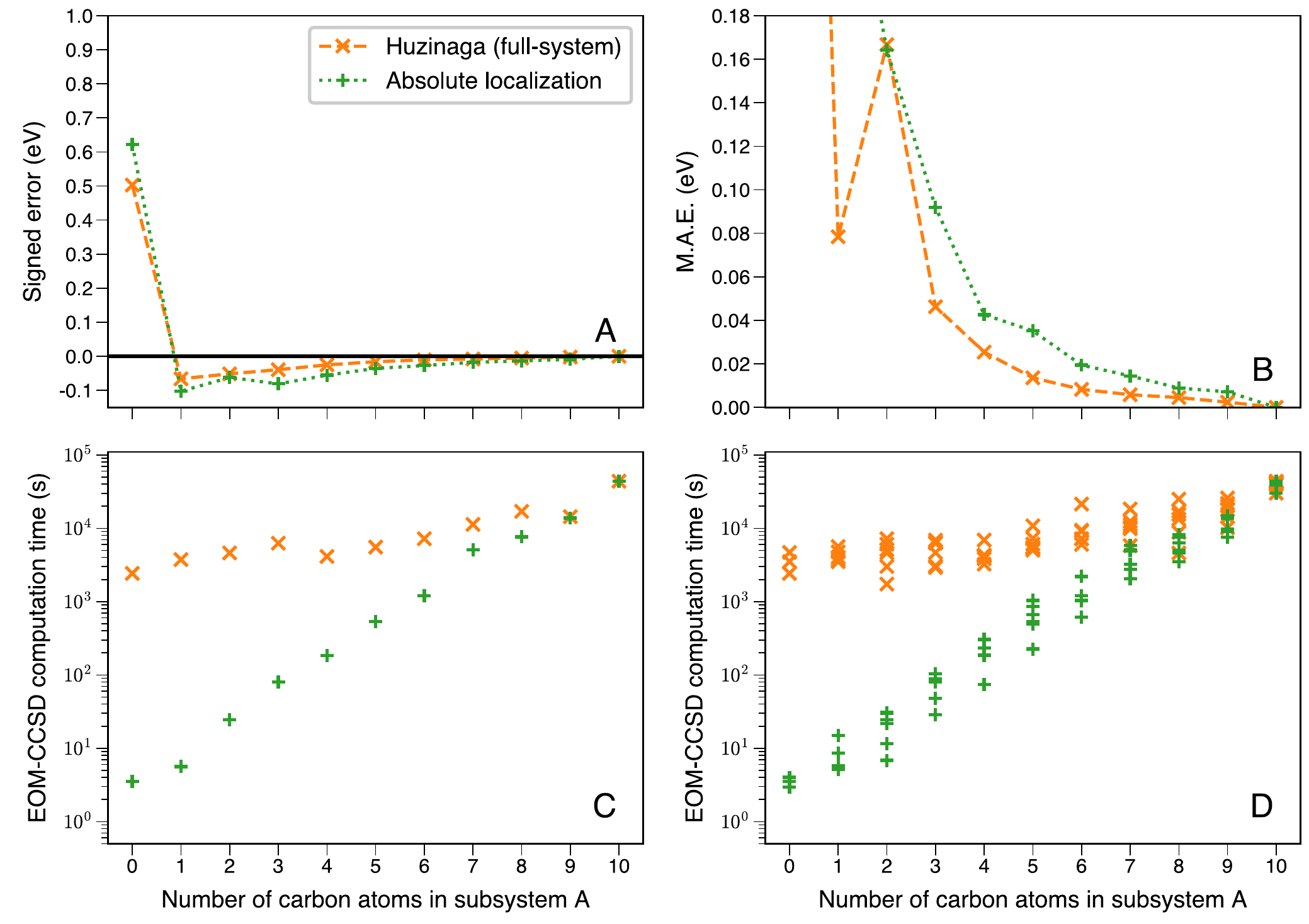}
    \caption{Embedded EOM-CCSD-in-M06 results as a function of increasing size of the subsystem A, showing: \textbf{(A)} Unsigned errors of vertical excitation energy for the first excited state of decanol with respect to canonical EOM-CCSD on the full system; \textbf{(B)} Mean absolute errors (M.A.E.) on the vertical excitation energies of 16 localized excited states; \textbf{(C)} EOM-CCSD computation time (on 16 cores of Intel Haswell E5-2680v3 processors) for the first excitation of decanol; \textbf{(D)} Average EOM-CCSD computation time over 16 localized excited states. 
    The x-axis indicates the number of carbon atoms included in subsystem A as indicated in Fig~\ref{fig:NTOs}. The NTOs of the 16 excitations examined are shown in Fig.~\ref{fig:NTOs}. Results are shown for the Huzinaga operator in the full-system basis (orange cross), and the Huzinaga operator in the subsystem basis (i.e., the absolute localization method; green plus sign).}
    \label{fig:op_comparison}
\end{figure}

Our embedding results are shown in Figure~\ref{fig:op_comparison}. In panel A, we show the unsigned errors of the first excitation of decanol where the basis set for subsystem A includes all the basis functions of the system (full-system, orange) or only the basis functions associated with the atoms of subsystems A (absolute localization, green).  In panel B, we show the mean absolute errors (M.A.E.) of all 16 local excitations mentioned in Figure~ \ref{fig:NTOs}, the values for all excitations are provided in the Supporting Information Table~S1 and Table~S2. 
%We observe that the errors of $\mu$ operator are always positive while Huzinaga operator could generate both positive and negative errors. This is expected as the $\mu$ operator shifts the overlapped orbitals to a very high energy level while Huzinaga operator flips the sign of these orbital energies. Therefore, the shifted orbitals by $\mu$ operators become inaccessible virtual orbitals in the subsequent excited-stated calculations while these by Huzinaga operators are accessible. The differences between the $\mu$ and Huzinaga operators are negligible once we consider a sufficiently large active subsystem; this is usually at three or more carbons included in the active system. The absolute localization method tends to a less accurate, but not significantly so.
Chemical accuracy is reached --- that is, less than 1 kcal/mol or 0.04 eV --- with four embedded carbons for the absolute localization method, compared to only three carbons for the full-system basis. 
This is because, for the HONTO $\rightarrow$ LUNTO transitions shown in Figure~\ref{fig:NTOs}, the LUNTOs are generally more delocalized than the HONTOs.
For the embedding calculations that use the full-system basis, the embedded EOM-CCSD uses the virtual orbital space of the full system, while the absolute localization strategy uses the virtual orbital space of the high-level subsystem only. Although both strategies have the same size of occupied space, the larger virtual space considered in the full basis strategy makes the results slightly more accurate. We want to emphasize here that absolute localization\cite{Chulhai2017,Chulhai2018} is not simply an extreme case of basis set truncation,\cite{Barnes2013,Bennie2015-ur,Bennie2016-tq} as both subsystems A and B are also solved self-consistently in the absolute localized basis at the DFT-in-DFT level, similar to ``freeze-and-thaw'' subsystem DFT methods.\cite{Jacob2006-hx,Jacob2008-bq,Gomes2008-sd,Gomes2012-pa}

The computation time of the EOM-CCSD calculations, the most time-consuming step in each of these methods, are shown in Figure~\ref{fig:op_comparison}C and ~\ref{fig:op_comparison}D. Although the absolute localization strategy is marginally less accurate, the subsequent embedded EOM-CCSD calculation are still 1--3 orders of magnitude faster due to the significantly reduced number of virtual orbitals.

Typically, $N^\text{A}_\text{vir} \leq N^\text{trunc}_\text{vir} < N_\text{vir}$, where $N^\text{A}_\text{vir}$ is the number of virtual orbitals for subsystem A in the absolute localized basis. This significant reduction in the number of virtual orbitals results in a more efficient EOM-CCSD calculation since EOM-CCSD scales as $N^4$ with respect to the number of virtual orbitals, compared to only $N^2$ with respect to the number of occupied orbitals.
A strategy to reduce the number of virtual orbitals in a correlated wave function treatment of excited states is crucial to cost saving, especially where large basis sets are required.
For example, Silva-Junior and co-workers found that at least a TZVP basis set are required for low-lying valence excited states when using correlated wave function methods,\cite{Schreiber2008,Silva-Junior2010a,Silva-Junior2010} while additional diffuse functions should be used for Rydberg states.\cite{Schreiber2008} In situations where such large basis sets are required, the speedup of our absolute localization embedding strategy would be even more significant than the cases shown in this subsection, which used cc-pVDZ basis set.  Given the significant computational cost savings with a relatively minor increase in error, we will use the absolute localization embedding strategy in the remainder of this paper.

\subsection{Eliminating Spurious Charge-Transfer States in TDDFT}\label{subsec:adenine} %--------

It is a well-known that TDDFT, particularly for local and semi-local exchange-correlation (XC) functionals, may result in artificial low-lying charge-transfer (CT) excited states.\cite{Magyar2007-bv,Tsuneda2014-nx,Dreuw2004}
This is caused by the incorrect 1/R asymptotic behavior of approximate XC kernels.\cite{Dreuw2003-bp,Dreuw2004}
%These states may be correctly simulated by correlated wave function methods, like EOM-CCSD\cite{Benda2016} \xw{Mayne I need to find a better example} and CASPT2\cite{Olaso-Gonzalez2009}. 
Long-range corrected functionals---such as CAM-B3LYP\cite{Yanai2004}, LRC-$\omega$PBEh\cite{Henderson2009}, LRC-$\omega$PBE\cite{Lange2008}, and $\omega$B97XD\cite{Chai2008}---correct for this issue by enforcing the correct 1/R asymptotic behavior at large distances.
Embedded TDDFT\cite{Casida2004-ys,Neugebauer2007-zd,Gomes2008-sd,Gomes2012-pa} is uniquely suited to solve this problem. Using the subsystem basis restricts the electron density of each subsystem on their own spatial regime, and therefore forbids the charge-transfer excitations between subsystems.
As a result, the artificially low long-range CT excitations between different weakly interacting subsystems will be eliminated.
Similarly, absolute localization can eliminate the spurious low-lying CT states since the subsystems are described with the subsystem basis. However, global and real CT electronic excitations will also be eliminated for the same reason.

\begin{figure}[!ht]
   \includegraphics[width=1.0\linewidth]{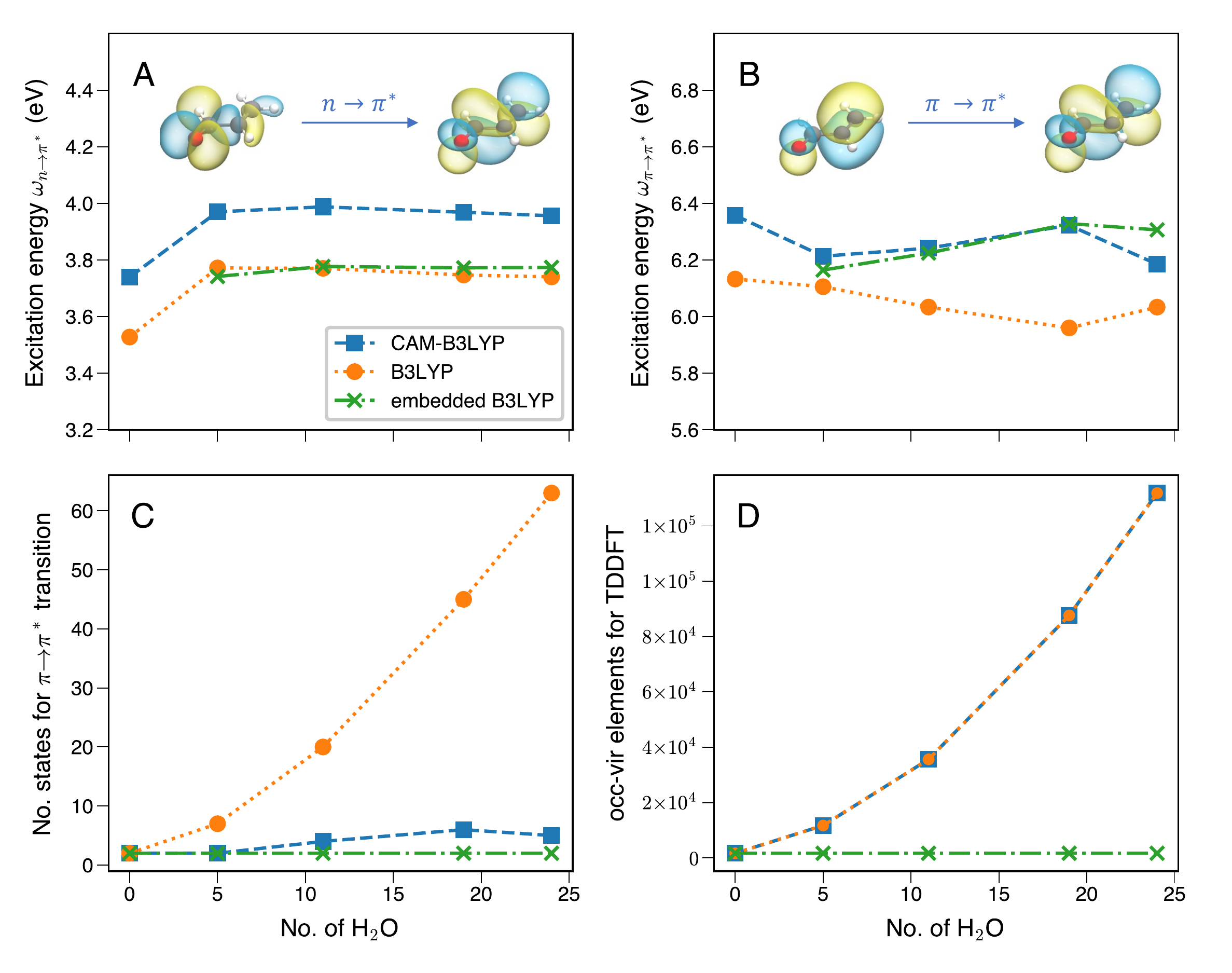}
   \caption{Vertical excitation energies in eV for the lowest singlet $n \rightarrow \pi^\text{*}$ and the brightest $\pi\rightarrow \pi^\text{*}$ transition of $\textit{s-trans}$ acrolein surrounded by different amount of water molecules. CAM-B3LYP (blue square), B3LYP (orange dot) and embedded B3LYP (green cross) are used. (A) the $n \rightarrow \pi^\text{*}$ transition, (B)the $\pi \rightarrow \pi^\text{*}$ transition, and (C) the number of states to be calculated to obtain the bright $\pi \rightarrow \pi^\text{*}$ transition, (D) the occupied-virtual orbital elements for the TDDFT calculations. The excitation energy, oscillator strength, HONTOs, and LUNTOs of all calculations are given in Figure S3 - Figure S5.}
   \label{fig:charge transfer}
\end{figure}

Figure~\ref{fig:charge transfer} shows the results for the lowest $n\rightarrow\pi^*$ and $\pi\rightarrow\pi^*$ transitions for $\textit{s-trans}$ acrolein surrounded by different amounts of water molecules.
In Figure~\ref{fig:charge transfer} (A), we observe that embedded B3LYP gives similar results to B3LYP for the $n \rightarrow \pi^\text{*}$ transition, while CAM-B3LYP predicts higher excitation energies.
In Figure~\ref{fig:charge transfer} (B), both embedded B3LYP and CAM-B3LYP predict higher excitation energies for the $\pi \rightarrow \pi^\text{*}$ transition.
In Figure~\ref{fig:charge transfer} (C), we give the number of excited states to be calculated to obtain the bright $\pi\rightarrow \pi^\text{*}$ transition.
For gas-phase acrolein, the first excited state is $n \rightarrow \pi^\text{*}$ transition, and the second excited state is a bright $\pi\rightarrow \pi^\text{*}$ transition responsible for the experimental absorption peak.
As we include more water molecules into the B3LYP and CAM-B3LYP calculations, the number of excited states that needs to be calculated before we obtain the bright $\pi\rightarrow \pi^\text{*}$ transition increases due to spurious low-lying CT states between the water molecules and arcolein; this was also observed in other similar systems, for example in uracil solution.\cite{Lange2008}
The spurious CT problem is severe in B3LYP, is largely mitigated by CAM-B3LYP, and is completely avoided by embedded B3LYP.
Figure~\ref{fig:charge transfer} (D) lists the product of the numbers of the occupied and virtual orbitals $N_{occ-vir}$, where the scaling of TDDFT calculation is $\mathscr{O}(N^3_{occ-vir})$. 

Based on the NTO analysis in Figure~S4 and S5, the $n \rightarrow \pi^\text{*}$ transition is localized with both B3LYP and CAM-B3LYP.
For this reason, embedded B3LYP gives similar result to B3LYP.
As we will further explore in Section~\ref{subsec:acrolein}, the environmental response is also weak; therefore, the ground-state embedding potential provides an accurate description for the environmental polarization resulting the close agreement between B3LYP and embedded B3LYP.
The CAM-B3LYP predict higher excitation energy than B3LYP because it has larger HF exchange at the long distance.\cite{Laurent2013}
The NTO analysis in Figure~S6 shows that the $\pi \rightarrow \pi^\text{*}$ transition has a small delocalized character with CAM-B3LYP and large delocalization with B3LYP.
CAM-B3LYP results in higher excitation energies than B3LYP due to the larger HF exchange at long distances, while embedded B3LYP increases the excitation energies (as compared to B3LYP) by forcing the $\pi \rightarrow \pi^\text{*}$ transition to be localized on the acrolein molecule, which increases the spatial overlap between HONTO and LUNTO\cite{Kuritz2011}.
As seen in Figures S3--S5, the NTOs of embedded B3LYP are fully localized on acrolein, and the calcualted oscillator strengths are much larger than those of B3LYP.

The example of acrolein in water suggests that absolute localization can simultaneously treat local excitation accurately, eliminate the spurious low-lying charge-transfer excitation effectively, and speed up the calculation significantly. In addition, it enforces the localization of $\pi \rightarrow \pi^\text{*}$ that are fictitiously delocalized by B3LYP.  Therefore, embedded TDDFT appears to be an excellent tool to study the excited states of solvated molecules.

%Laurent and Jacquemin\cite{Laurent2013} reviewed TDDFT benchmarks and found that traditional global hybrid functionals provide rather satisfactory results for localized valence excited states. In these cases, using embedded TDDFT is able to persevere the good quality of the local valence states and improve the quality of fictitiously delocalized $\pi \rightarrow \pi^\text{*}$  states.

%However, it should not be used when the excited states of interest are global and/or are real CT states, and it does not improve the description of Rydberg states.
%\dvcc{This is the first comment about Rydberg states in this section, I think. Is it needed? I don't know what this means}

\subsection{Accounting for Environment Polarization} \label{subsec:acrolein} %----------------

%\dvcc{Paragraph: Why is polarization response important}

For excited states calculations, there are two types of polarization to consider: (1) The polarization of the environment to the ground-state of the high-level subsystem A; and (2) the polarization response of the environment to excitations in the high-level subsystem A. The former is exactly captured within projection-based embedding methods, at least when using the full system basis. For the latter in a quantum embedding framework, there has so far been three general strategies:
(i) accounting for the coupled response of the environment within a TDDFT framework;\cite{Neugebauer2007-zd,Hofener2012-mb,Chulhai2016-wg}
(ii) including important environment orbitals in the description of the high-level subsystem A within correlated wave function methods like EOM-CCSD;\cite{Bennie2017} or
(iii) using a state specific embedding potential.\cite{Khait2010-hn,Daday2013-kd,Wesolowski2014,Ricardi2018}
Within the embedded TDDFT framework, the response of the environment is often accounted for using a coupled frozen-density embedding term added to the TDDFT kernel.\cite{Neugebauer2007-zd} This coupled response approach have also been extended to projection-based TDDFT embedding with the $\mu$ projection operator.\cite{Chulhai2016-wg} Within conventional projection-based excited states embedding,\cite{Ding2017,Bennie2017,DeLimaBatista2017} all of the environment virtual orbitals, unless some basis set truncation scheme\cite{Barnes2013,Bennie2015-ur,Bennie2016-tq} is used, are included in the subsequent correlated WF calculation of the high-level subsystem. The polarization response of the environment therefore only requires inclusion of selected important environment occupied orbitals into the high-level subsystem.\cite{Bennie2017} However, as far as we are aware, there is no \textit{a priori} method to determine which environment orbitals are important, and current methods\cite{Bennie2017} rely on a guess-and-check scheme.%\dvcc{VERIFY/TECHNICALLY, THEY USE CIS TO DETERMINE IMPORTANT ORBITALS} \xw{The approach to determine which orbitals of the environment to be included was as follows: (i) HF on the whole system; (ii)Selecttheatoms of the photoactive molecule (or part of the molecule). (iii) Run CIS (or LR-TDDFT, CC2) for subsystem A; (iv)Increase the number of orbitals in the embedding region by 1 and check if the excitation energy is larger than some threshold value (e.g 0.01 eV); (v) Add all those orbitals which cause the largest improvement to the CIS and re-run with eEOM-CCSD.}

\begin{figure}[ht]
   \includegraphics[width=0.6\linewidth]{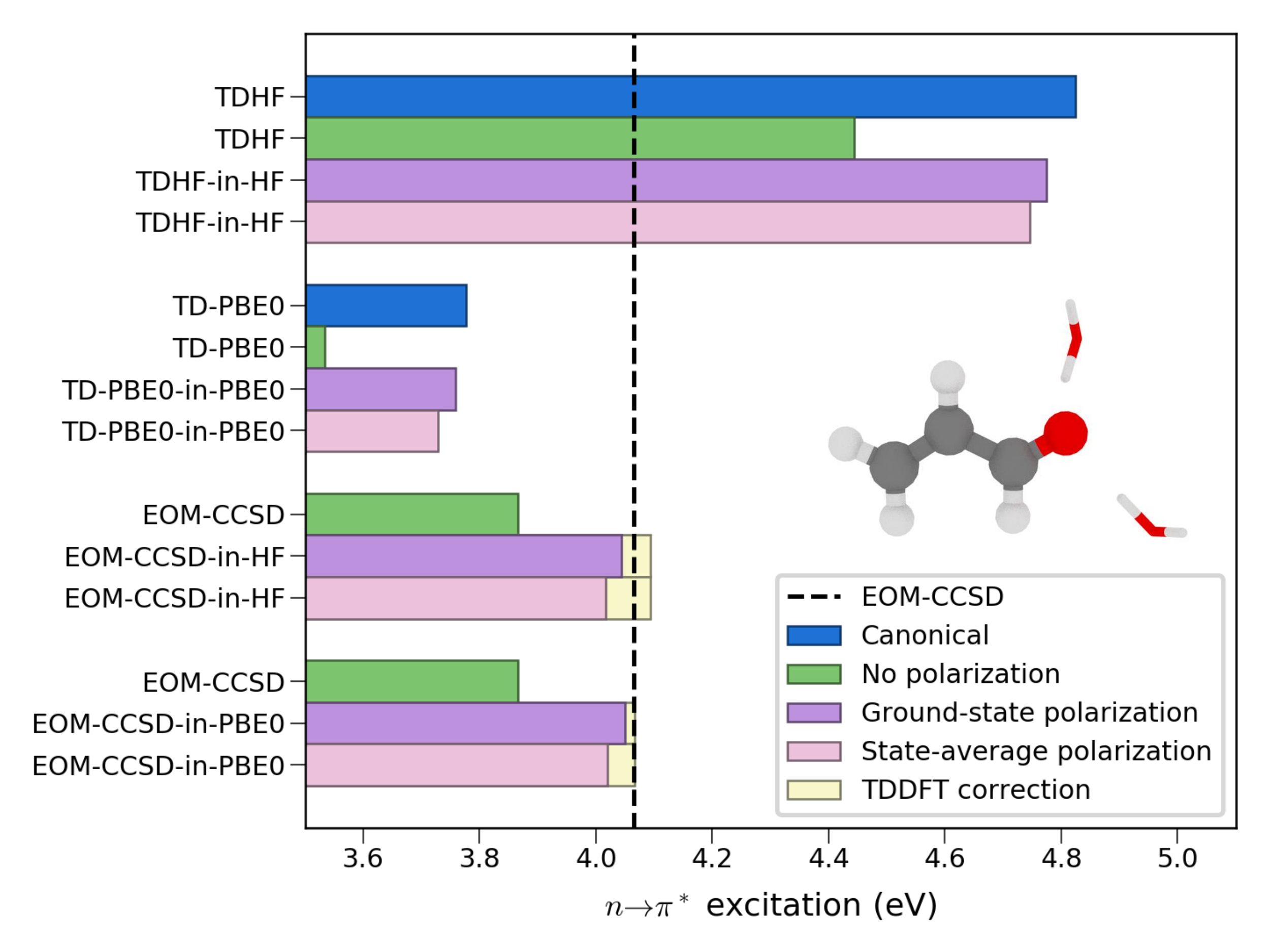}
   \caption{Excitation energies in eV of the $S_0 \rightarrow S_1$ ($n\to\pi^*$) transition for acrolein+2water (inset). Results are shown for canonical TDHF/TDDFT on the full system (blue); canonical TDHF/TDDFT/EOM-CCSD on acrolein only, with no polarization from water molecules (green); Embedded TDHF/TDDFT/EOM-CCSD with ground-state (purple) and state-average (pink) polarization; and the TDHF/TDDFT polarization correction (light yellow). The inset shows the system with acrolein included in subsystem A, and the two water molecules included in subsystem B; these are represented by grey (carbon), red (oxygen), and white (hydrogen).}
   \label{fig:acrolein}
\end{figure}

We examine the effects of various types of polarization on the lowest $n\to\pi^*$ transition in an acrolein+2water system, which was previously studied using $\mu$ projection-based excited states embedding.\cite{Bennie2017} In Figure~\ref{fig:acrolein},
We will first consider the effects of ground-state polarization on this transition. For embedded TD-HF-in-HF (or TD-PBE0-in-PBE0), we find that ground-state polarization (purple bars) --- that is, absolute localization embedding without consideration for the environment polarization response --- already accounts for $\sim90\%$ of the missing polarization in the canonical full system results (blue bars) when compared to the isolated acrolein results (``no polarization;'' green bars).  For embedded EOM-CCSD-in-HF(or PBE0), ground-state polarization similarly captures $\sim90\%$ of the polarization, regardless of the quality of the low-level method (HF or PBE0 in the case). We will discuss the effect of the lower level method in the next subsection. The small fraction of  missing polarization --- around $10\%$ in the case of the hydrogen-bonded acrolein+2water examined here --- is due to a missing environment response. This is because there are no orbitals, neither occupied nor virtual, from environment that are included in the high-level subsystem A in the absolute localization method.

%Comparing embedding to canonical results shows that the quality of embedding results is mainly determined by the quality of method for treating excited states and merely affected by choice for the ground state. Besides, the relaxed scheme always brings the excitation energy down instead of getting closer to the canonical result, as we originally expected. Here is the possible explanation for the unexpected behavior of relaxed scheme: like the widely-used solvation model polarizable continuum model (PCM)\cite{Aguilar1993, Cammi1995, Mennucci1998}, which could be considered as an implicit and less-refined embedding method.

In order to include the polarization response of the environment, we will use two approaches: (1) Using a state-averaged approach; and (2) Using a TDDFT correction. In the state-averaged approach (pink bars in Figure~\ref{fig:acrolein}), subsystem B is polarized to the average density of the ground-state and first excited state of subsystem A. However, we find that this approach lowers the excitation energy, which in this case leads to increased error.    
%\xw{Currently, ground and excited state of A are destabilized, the ground state of B is stabilized. Doing: calculate the supermolecular (A+B) to check whether the variational principle is preserved. Checked: DFT-in-DFT energy decreases, variational principle is reserved.}
Possible reasons for the behaviour of state-average embedding potential are:
1. Huzinaga operator shifts the occupied orbitals of the environment to positive energy levels, but does not removed them, and therefore creates fictitious virtual orbitals that might be involved in the subsequent excited-state calculations. 
%When ground-state embedding potential is used, the errors introduced by Huzinaga operator is not significant, since Huzinaga operator achieves the similar accuracy with $\mu$-operator as shown in Fig~\ref{fig:op_comparison}. 
This unrealistic relaxation on excitation energies might be exaggerated by the optimizing the subsystem densities using state-average embedding potential self-consistently. 
2. The errors associated with embedding are more complex than simply the lack of environmental response.  Recent work has shown that the accuracy of different embedding strategies for excitation energies are largely similar with or without environmental response.\cite{Ricardi2018}  It is possible that there are errors associated with the embedding potential that get exaggerated when the state-averaged embedding potential is used; therefore, although the electrostatic potential is more accurate due to environmental response, the overall potential is of similar quality.  Regardless of the exact reason, this suggests that including the response of the environment does not always lead to better accuracy.

The second approach to account for the environment polarization response is by using a correction term obtained by TDDFT; this correction is defined in eqn.~\ref{eq:corrected_wf_energy}. We explore a similar correction term for ground state energies in ref.~\citenum{Chulhai2017}, where the correction accounted for the errors introduced by using the absolutely localized basis --- errors introduced, for example, by preventing electrons from moving between subsystems. Analogously, by performing an embedded excitation calculation of subsystem A in the absolutely localized basis, we lose the contributions to the excitation from coupling to subsystem B. In ref.~\citenum{Bennie2017}, this missing coupling is described in terms of orbital transitions --- transitions involving both occupied and virtual orbitals of subsystem B are missing --- while in ref.~\citenum{Chulhai2016-wg}, this missing coupling is described in terms of a coupled environment response --- changes to the density of subsystem B when subsystem A is excited are missing. The correction presented in eqn.~\ref{eq:corrected_wf_energy} corrects for this missing coupling (or environment polarization response) using the canonical TDDFT excitation on the full system. Since we are interested in cases where canonical EOM-CCSD on the full system is prohibitively expensive and the embedded EOM-CCSD is the most time consuming step of our methodology, there are systems in which a full system TDDFT calculation is an insignificant cost to pay to include the environment response; of course, for some systems this cost could become intractable.  

The results with this TDDFT correction are also shown in Figure~\ref{fig:acrolein} and included in the Supporting Information Table~S3. We observe that this correction result in similar accuracy, regardless of whether we add the correction to the ground-state polarization or state-average polarization approaches. With this correction, we also observe a small dependence on the quality of the lower level method, with the TDHF corrected EOM-CCSD-in-HF results 0.03 eV higher than the benchmark canonical EOM-CCSD excitation energy, while the TDDFT corrected EOM-CCSD-in-DFT results exactly, errors less than 0.01 eV, reproduced the benchmark. These corrected errors are small, given that canonical TDHF and TD-PBE0 have errors of 0.76 and 0.30 eV, respectively, compared to canonical EOM-CCSD. Our results are similar in accuracy to the those in ref.\citenum{Bennie2017} for a smaller embedded EOM-CCSD orbital space. 
%and the additional cost of a canonical TDDFT on the full system.
We will continue to examine the accuracy of this correction approach and its dependence on the quality of the lower level method in next subsection.

In summary, our results suggest that ground-state polarization of the environment provides adequate accuracy.  Additional schemes to include the polarization response of the environment do not always increase the accuracy.  However, the error associated with ground-state polarization for EOM-CCSD-in-PBE0 or EOM-CCSD-in-HF is only 0.015 eV and 0.019 eV, respectively.  Therefore, this absolute localization scheme with ground-state polarization is very accurate for WF-in-DFT embedding, and the lack of need of polarization response of the environment means the algorithm remains highly computationally efficient.  

\subsection{Quality of Low Level Method} \label{subsec:7HQ} %-----------------------------------

%\begin{figure}[!ht]
%   \includegraphics[width=1.0\linewidth]{figures/7HQ.eps}
%   \caption{Solvent-induced vertical excitation energy shifts $\Delta\omega_{\pi\rightarrow\pi^*}$ for the lowest $\pi\rightarrow\pi^*$ transition in cis-7HQ chromophore. aug-cc-pVDZ basis set is used for all calculations. LDA\cite{Vosko1980} is used for DFT and TDDFT calculations. Absolute localization is used for embedding scheme. The detailed wavenumbers and relaxed data are given in Table~S2 of the Supporting Information}
%       \label{fig:7HQ}
%\end{figure}

\begin{figure}
    \includegraphics{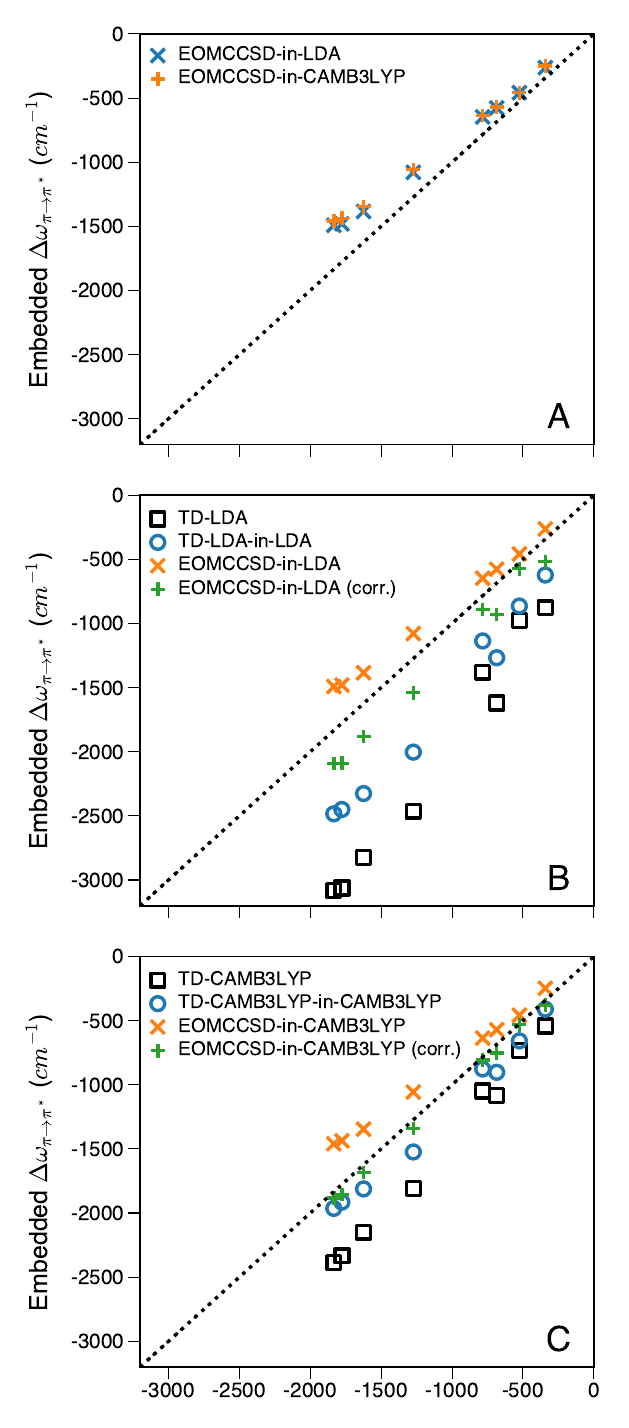}
    \caption{Solvation energy shifts of the lowest $\pi\to\pi^*$ excitation in \textit{cis}-7-hydroxyquinoline ($\Delta\omega_{\pi\to\pi^*}$) compared to the canonical EOM-CCSD results. (A) Embedded EOM-CCSD-in-LDA and EOM-CCSD-in-CAMB3LYP with ground-state polarization only. (B--C) Comparison of canonical TDDFT, embedded TDDFT (ground-state polarization), and embedded EOM-CCSD with ground-state and TDDFT corrected (corr.) polarizations, using (B) LDA and (C) CAMB3LYP as the low level exchange-correlation functionals.}
    \label{fig:7HQ}
\end{figure}

To examine the effects of the choice exchange-correlation functional on the quality of the EOM-CCSD-in-DFT embedding, we looked at the solvent shifts of \textit{cis}-7-hydroxyquinoline (\textit{cis}-7HQ) due to eight hydrogen-bonded complexes: water, ammonia, methanol, formic acid, two water molecules, ammonia-water-water trimer, ammonia-water-ammonia trimer, and ammonia-ammonia-water trimer; solvent shifts are measured with respect to the gas phase 7HQ. These results are shown in Figure~\ref{fig:7HQ} (and tabulated in Supporting Information Table S4), where we plot the solvation energy shifts of the lowest $\pi\to\pi^*$ excitation ($\Delta\omega_{\pi\to\pi^*}$) compared to the canonical EOM-CCSD results.
In this figure, we examine two exchange-correlation functionals, LDA and CAM-B3LYP, and show the results for canonical TDDFT, embedded TDDFT, embedded EOM-CCSD with ground-state polarization, and embedded EOM-CCSD with ground-state polarization and the TDDFT correction in eqn.~\ref{eq:corrected_wf_energy}.

Of the methods explored, canonical TDDFT is the least accurate, with MAEs of 930 and 406 cm\textsuperscript{-1} for LDA and CAM-B3LYP, respectively. These errors also appear to be dependent on the size of the solvent molecule(s).
The solvent molecule(s) from smallest to largest solvation shifts (right to left in Figure~\ref{fig:7HQ}), are CH\textsubscript{3}OH, H\textsubscript{2}O, HCOOH, NH\textsubscript{3}, 2H\textsubscript{2}O, NH\textsubscript{3}-H\textsubscript{2}O-NH\textsubscript{3}, NH\textsubscript{3}-H\textsubscript{2}O-H\textsubscript{2}O, and NH\textsubscript{3}-NH\textsubscript{3}-H\textsubscript{2}O.
Fradelos and co-workers examined the same systems\cite{Fradelos2011} and suggested that the poor performance of TDDFT is most likely caused by the incorrect 1/R asymptotic behavior of the LDA potentials; they found that the statistical average of orbital potentials (SAOP)\cite{Gritsenko1999}, an asymptotically correct exchange-correlation potential, gives better results than the LDA and PW91 functionals in both embedded and canonical TDDFT calculations.
Therefore, this size-depedency of the canonical TDDFT errors can be attributed to erroneously delocalizing---and therefore lowering the energy of---the excited state over the entire system\cite{Grimme2003,SosaVazquez2015,Fradelos2011}.
Canonical TD-CAM-B3LYP, which has a factional 1/R dependence,\cite{Yanai2004} provides significantly more accurate excitations energies compared to TD-LDA.

Embedded TDDFT-in-DFT, with ground-state polarization only, performs better than canonical TDDFT, with MAEs of 539 and 153 cm\textsuperscript{-1} for LDA and CAM-B3LYP, respectively.  As explored in Figure~\ref{fig:charge transfer}, this is due to restricting the transition to the high-level subsystem only through absolute localization, and preventing the delocalization of the $\pi^*$ state over the full system. This restriction results in more accurate localized excitations, regardless of the asymptotic behavior of the exchange-correlation functionals. Therefore, for these types of localized solvated excitations, embedded TDDFT can be more accurate and significantly cheaper than canonical TDDFT.  

%\dvcs{Since we use absolute localization scheme for the TDDFT-in-DFT calculations, the TDDFT region is size-intensive in all embedding calculations. This might explain why TDDFT-in-DFT performs better than supermolecular TDDFT, where the TDDFT region is size-extensive and therefore introduces more errors.}

For the embedded EOM-CCSD-in-DFT calculations with ground-state polarization only, we find that the choice of low-level functional---either LDA or CAM-B3LYP, shown in Figure~\ref{fig:7HQ}A---does not affect the results significantly; this was also observed for acrolein in Figure~\ref{fig:acrolein}.
The EOM-CCSD-in-LDA and EOM-CCSD-in-CAMB3LYP have MAEs of 183 and 204 cm\textsuperscript{-1}, respectively.

Given that embedded EOM-CCSD-in-DFT calculations do not significantly depend on the choice of the DFT exchange-correlation functional, it suggests that the reason the embedded TDDFT-in-DFT calculations do depend on the exchange-correlation functional is due to the TDDFT functional and the not the environment DFT functional.  Therefore, it is not surprising that CAM-B3LYP provides significant better accuracy compared to LDA.  We can conclude that the low level functional does not significant influence our embedded excitation energies.  This is consistent with our previous results, which found that the choice of exchange-correlation functionals have a minimal effect on the ground state properties of absolutely localized embedded WF-in-DFT.\cite{Chulhai2017,Chulhai2018}

Including state-average polarization (results not shown) only marginally improves ($\sim 20$ cm\textsuperscript{-1}) on the ground-state polarization results. The additional cost of multiple embedded TDDFT-in-DFT calculations for the state-average approach therefore makes it less desirable. However, for the TDDFT correction in eqn.~\ref{eq:corrected_wf_energy}, we find that the results of EOM-CCSD-in-CAMB3LYP are significantly improved (MAE of 50 cm\textsuperscript{-1}), while the EOM-CCSD-in-LDA results remain the same (MAE of 209 cm\textsuperscript{-1}).
Because this correction depends on the canonical TDDFT on the full system, the corrected EOM-CCSD-in-DFT are therefore dependent on the quality of the XC functional used.
With this TDDFT correction, we also observe that there is no size-dependence in the solvation energy shifts, unlike what was observed for ground-state polarization EOM-CCSD-in-DFT.
We ascribe this to the larger polarization response of the larger solvation molecules; this response is not accounted for by the ground-state embedding but is accounted for by the TDDFT correction.

\section{A Systematically Improvable Method: Green Fluorescent Protein} \label{subsec:GFP} %---------------

%All embedding calculations were performed in pySCF\cite{Sun2018}. All canonical TDDFT calculations were performed in Gaussian 16\cite{g16}. 

In this subsection, we will show how the absolute localization method is systematically improvable by examining a 161-atom model of green fluorescent protein (GFP). The geometry is taken from ref~\citenum{Kaila2013}, which contains the neutral \textit{p}-hydroxybenzylidene-imidazolinone chromophore (chro), nine amino acid residues including T62, Q69, Q94, R96, H148, V150, T203, S205, and E222, and four water molecules.

\begin{figure}[!ht]
   \includegraphics[width=0.4\linewidth]{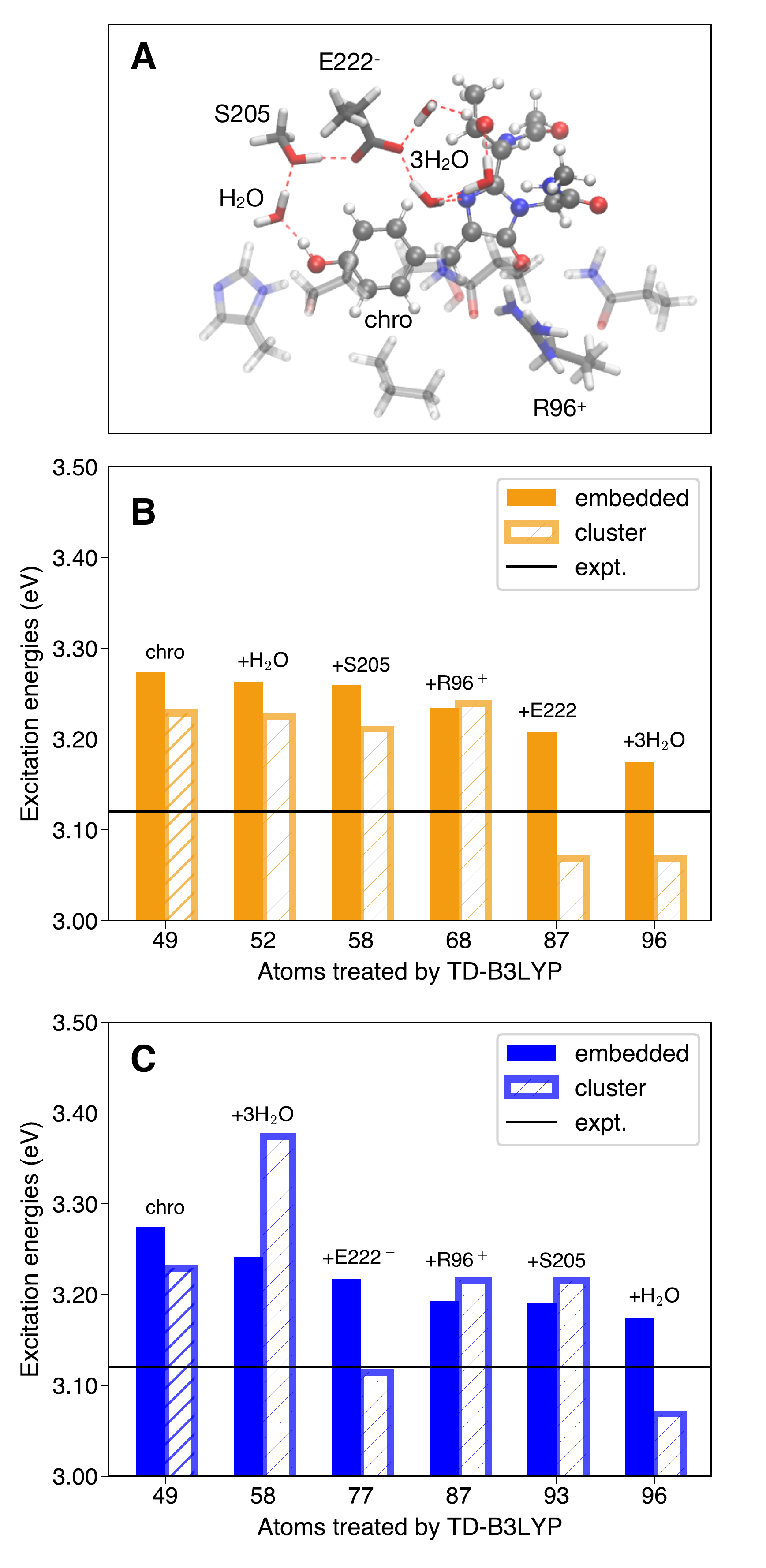}
   \caption{(A) The 161-atom model. The ball-and-stick model represents the chromophore (chro), which is included in all calculations. The tube model includes the hydrogen bonds network made by four water molecules, S205, E222$^-$, and its counter ion R96$^+$, which are added to the cluster and embedded calculations. The translucent tube model includes six amino acid residues which are always treated as the environment subsystem in embedded calculations. 
   (B-C) The excitation energy of the brightest excited state in cluster TDDFT and embedded TDDFT-in-DFT calculations by successively adding the residues in the hydrogen bond chain from (B) right to left and (C) left to right. The black line indicates the excitation energy 3.12 eV from TDDFT calculation on the full system.}
   \label{fig:GFP}
\end{figure}

We systematically increased the size of the subsystem A in two ways: (i) by successively adding the H\textsubscript{2}O, S205, E222\textsuperscript{-}/R96\textsuperscript{+}, and 3H\textsubscript{2}O residues to the chromophore; or (ii) by successively adding the same moieties but in the reverse order.
We performed two sets of calculations: (1) isolated TDDFT calculations (no embedding) where the size of the system included only the chromophore and additional moieties and (2) embedded TDDFT-in-DFT (ground-state embedding) where the high-level system was the same size as the isolated TDDFT calcualtions but embedded in the the remainder of the 161-atom model as the environment subsystem B.  We note that these calculations are of similar computational cost since the TDDFT cost is identical.
These results are shown in Figure~\ref{fig:GFP}, where Figure~\ref{fig:GFP}A shows the 161-atom model with the chromophore (balls and sticks model), residues that are successively added to the high-level subsystem (opaque tubes model), and residues in the environment (translucent tubes model). The transition of interest, shown in Figure~\ref{fig:GFP}B, was identified by examining
the calculated oscillator strengths and shapes of the natural transition orbital (NTOs).\cite{Martin2003}  Our canonical TDDFT using B3LYP on the full 161-atom model showed that the brightest low-lying state (oscillator strength \textit{f}=0.76 a.u.) was 3.12 eV, which is in good agreement with experimental results of 3.12-3.14 eV for the neutral A-form GFP.\cite{Brejc1997-ub,Creemers1999-kj,Creemers2000-xp}

The results of successively increasing the size of the high-level subsystem are shown in Figures~\ref{fig:GFP}C and D for both the cluster TDDFT and embedded TDDFT-in-DFT results. The black line is the canonical TDDFT results on the full 161-atom model.
For the cluster (no embedding) calculations, we observe that the results are erratic; increasing the size of the cluster does not always translate to a more accurate excitation energy.
However, for the embedded calculations, we always obtain a more accurate results with a larger subsystem A.
The fundamental difference between these two (cluster and embedded) models is the presence of the ground-state polarization of the environment.
For embedded TDDFT-in-DFT, the ground-state polarization --- the most important contribution to solvent shifts as explored in Figure~\ref{fig:7HQ} --- is always included, and adding residues to the high-level subsystem systematically increases the amount of the system which is responding to the excited state.
This systematic improvability of projection-based embedding was previously observed for ground state properties,\cite{Goodpaster2014,Chulhai2017} and we show here that it also applies to the excited states properties.

Additionally, embedding calculations allow for a systematic study of the relative importance of the water molecules and residues in the excitation energy.  A series of embedding calculations allow for the decoupling of the environment effects between ground state polarization and excited state polarization.  By including the entire system for all calculations, the environment polarization (in the ground state) is included and by expanding the size of the TDDFT calculation, one can then determine how much each neighboring water molecule or residue polarizes to the excited state.   This allows one to understand the role and relative importance of the environment to the excited state of the chromophore.  

Embedding calculations are robust irrespective of which other water molecules or residues are included in the excited state calculation.   For instance, starting with the chromophore (panel C), inclusion of the 3H$_2$O water molecules lowers the excitation energy by -0.032 eV.  For the pathway followed in panel B, the inclusion of the 3H$_2$O water molecules occurs last, and lowers the excitation energy by -0.033 eV, nearly identical.  Therefore, regardless of when that moiety is included, embedding calculations can be used to determine the importance of that moiety in excited state polarization.  

These results, and the results from Figure~\ref{fig:op_comparison}, demonstrate that systematic improvability is retained when using projection based embedding for excited states. Which is to say that larger subsystems for the excited state regions will lead to more accurate calculations.  Additionally, our calculations on GFP show that these calculations can be used to determine the relative importance of chemical moieties to the excitation energy in a robust manner.    

\section{Summary and Conclusions}%===========================================================

We explored the behavior of absolute localization projection-based embedding method for excited states. We show that, for localized excited states with embedded EOM-CCSD, the absolutely localized method is marginally less accurate but significantly cheaper than embedding using the basis set for the full system.  Additionally, we show how this method can accurately eliminate spurious low-lying excited states in embedded TDDFT.
We found that the ground-state polarization of the environment, naturally included with embedded excited states calculations, accounts for $\sim90\%$ of the excitation energy shifts due to solvents. The remainder was ascribed to environment polarization response.  We proposed two approaches to include this polarization response --- a state-average approach and a TDDFT correction approach --- and found that the TDDFT correction accurately accounts for the environment response; this TDDFT correction is only applicable to embedded EOM-CCSD, where the full system TDDFT cost is small compared to the cost of the subsequent EOM-CCSD calculation. We showed that this embedding strategy is systematically improvable, and we are always guaranteed more accurate results for larger regions included in the high-level subsystem.  Finally, we showed that embedding could be a useful tool for understanding the relative importance of chemical moieties to excited states.  

Our results show correlated wave function level of accuracy at a significantly reduced cost. Absolutely localized basis\cite{Chulhai2017} reduces computational cost by aggressively truncating the basis functions to only those on atoms involved in the excitation. This truncation results in far fewer virtual orbitals, and thus lower computational cost.  We used EOM-CCSD as the excited states wave function method throughout this paper, however, we see no limitations in applying this absolutely localization embedding strategy to other excited states wave function methods as well, like the complete active space with second-order perturbation theory (CASPT2).\cite{Andersson1990-qk,Andersson1992-zo} We are currently exploring such embedding strategies.

\begin{acknowledgement}
                                                                                      
This research was carried out within the Nanoporous Materials Genome Center, which is supported by the U.S. Department of Energy, Office of Basic Energy Sciences, Division of Chemical Sciences, Geosciences, and Biosciences under Award DE-FG02-17ER16362. The authors acknowledge the Minnesota Supercomputing Institute (MSI) at the University of Minnesota and the National Energy Research Scientific Computing Center (NERSC), a DOE Office of Science User Facility supported by the Office of Science of the U.S. Department of Energy under Contract No. DE-AC02-05CH11231, for providing resources that contributed to the results reported within this paper.
\end{acknowledgement}

\begin{suppinfo}
%\dvcc{Make sure this is up to date.}
%Embedding results using Huzinaga operator/monomer basis, Huzinaga operator/supermolecular basis, and $\mu$ operator/supermolecular basis for decanoic acid, decene, decanal, decanamine, chlorodecane and decanol; 
NTO analysis and embedding results for the third excited states of chlorodecane and decanol; numerical data for Figure 3; The NTO analysis for acrolein in different number of water molecules with embedded B3LYP, canonical B3LYP, and CAM-B3LYP discussed in Figure \ref{fig:charge transfer}; numerical data for $\Delta\omega_{\pi\rightarrow\pi^*}$ discussed in Figure \ref{fig:7HQ}; geometries used in this paper.
\end{suppinfo}

\bibliography{references}

\end{document}